# Repurposing the Combination Drug of Favipiravir, Hydroxychloroquine and Oseltamivir as a Potential Inhibitor against SARS-CoV-2: A Computational Study

Pooja and Papia Chowdhury [*]

Department of Physics and Materials Science & Engineering, Jaypee Institute of Information Technology, Noida 201309, Uttar Pradesh, India.

## Abstract

The virus SARS-CoV-2 has created a situation of global emergency all over the world from the last few months. We are witnessing a helpless situation due to COVID-19 as no vaccine or drug is effective against the disease. In the present study, we have tested the applicability of some combination drugs against COVID-19. We have tried to understand the mechanism of action of some repurposed drugs: Favipiravir (F), Hydroxychloroquine (H) and Oseltamivir (O). The ADME analysis have suggested strong inhibitory possibility of F, H, O combination towards receptor protein of 3CL$^{pro}$ of SARS-CoV-2 virus. The strong binding affinity, number of hydrogen bond interaction between inhibitor, receptor and lower inhibition constant computed from molecular docking validated the better complexation possibility of F+H+O: 3CL$^{pro}$ combination. Various thermodynamical output from Molecular dynamics (MD) simulations like potential energy ($E_g$), temperature (T), density, pressure, SASA energy, interaction energies, Gibbs free energy ($\Delta G_{bind}$) etc., also favored the complexation between F+H+O and CoV-2 protease. Our In-Silico results have recommended the strong candidature of combination drugs Favipiravir, Hydroxychloroquine and Oseltamivir as a potential lead inhibitor for targeting SARS-CoV-2 infections.

## 1. Introduction

In 21st century, from the last few months the whole world is witnessing the pandemic due to the recent outbreak by the disease COVID-19 caused by novel coronavirus. It is believed that the virus was originated from the wet meat market of Wuhan city of China sometime in December in 2019 [Wang, Hu, et al., 2020; Guan et al., 2020]. Corona viruses are not new to the mankind. From the last few centuries, mankind had perceived the presence of these viruses in the form of avian flu virus around 2003 [Keil, et al., 2006], Severe Acute Respiratory Syndrome Coronavirus (SARS-CoV) around 2003 [Rota, et al., 2003], Middle East Respiratory Syndrome Coronavirus (MERS-CoV) around 2012 [Su, et al., 2016]. The latest pandemic disease COVID-19 that has created a public health emergency in this series aroused due to novel coronavirus: SARS-CoV-2 [Huang, et al., 2020]. The whole scientific fraternity in the world is working hard day and night to get a medicine/vaccine to fight against this deadly virus, but the reality is that till date we have no definite medication for COVID-19. As of today, the 5th November, 2020 according to WHO data there are 47,930,397 confirmed patients of COVID-19 disease and 1,221,781 deaths worldwide have been reported [https://www.who.int]. The transmission of the disease as well as the death toll due to COVID-19 expanding exponentially day by day. CoV-2 can easily spread by contact transmission. The infection can easily be transmitted through respiratory droplets and also through surface contamination. Aerosol transmission may be considered as another way of transmission of the disease [Rothe, et al., 2020]. Till now research outputs says that the virus can easily be transmitted by symptomatic asymptomatic and presymptomatic patients [Furukawa, et al., 2020; Chunyang, et al., 2020; Ghinai, et al., 2020]. There is an urgent need of effective treatment method to limit the transmission of this disease as early as possible.

Coronavirus is spherical in shape having diameter between 80 – 160 mm. The spherical envelope surface is covered with spike glycoproteins (S), membrane proteins (M) and envelope proteins (E) [Woo, et al., 2005; Gorden, et al., 2020]. The main envelope of virus contains a spiral nucleocapsid which is formed by genomic RNA and phosphorylated nucleocapsid (N) protein [Priyadarsini, et al., 2020]. The main genome of CoV is comprised of a longest known genome among RNA viruses which is a single-stranded positive-strand RNA ranging from 26 Kb to 32 Kb in length. Coronaviruses can be divided into four categories: α, β, γ, and δ. α and β coronaviruses only infect mammals [Heidary, et al., 2020], while γ and δ mainly infect birds [Muradrasoli, et al., 2010]. SARS-CoV-2 is a novel β-coronavirus. The S proteins of the virus initiate the attachment and entry to the host cells through the receptor binding domain (RBD) which is loosely attract and attach to the virus surface [Jin, et al., 2020]. All corona viruses use some key receptors to enter the host human cells. For SARS-CoV-2, the key receptor is angiotensin converting enzyme 2 (ACE2) [Velavan, et al., 2020]. After entering the host cell, airway trypsin-like protease (HAT), cathepsins and transmembrane protease serine 2 (TMPRSS2) split the S proteins of the virus and establish the penetration changes. Recent medical reports suggested that for CoV-2, the single N501T mutation in spike proteins may definitely have enhanced the virus's binding affinity for ACE2 [Ortega, et al., 2020]. So, development of targeted spike glycoprotein therapeutics against SARS-CoV-2 will definitely be a suitable option to combat COVID-19. In this direction many vaccines, antiviral drugs, broad spectrum antibiotics, herbal medicines nebulization techniques have been in use to reduce the viral load for the affected patients [Hendaus, 2020; Chowdhury, 2020; Khan, et al., 2020].

Researchers around the world are working to develop around 170 types of vaccines to deal with corona viruses out of which around 31 vaccines are following various human trial phases [Chang et al., 2016; Grein et al., 2020]. Though according to the medical application protocol, vaccine typically requires years of research, testing and several phase wise human trials before reaching to the clinic, but due to our latest health emergency scientists are racing to provide an effective and efficient vaccine/drug for the mankind as soon as possible. Similarly, drug development also is a time-consuming process which includes laboratory development, animal testing, and clinical trials in people before it reaches to market which can take a decade for a new drug to appear in market. So, to combat with current COVID emergency, researchers and medical practitioners are working on different available drugs that are already approved for other medical conditions, or have been tested on other viral diseases. The concept is known as drug repurposing which has become very important way now to identify potential drugs against the CoV-2 virus [Singh et al., 2020]. Among different options of repurposing drugs, mainly antiviral drugs which are already in use for other viral diseases like influenza, Ebola, HIV, hepatitis, Zika etc., are largely been used to treat the COVID-19 disease [Suryapad et al., 2020; Manoj et al., 2020; Chowdhury, 2020]. Usually any antiviral drug targets the infecting virus in three vital stages. The drug may resist the virus from entering the living cell, it can prevent the virus from replicating inside the human organ or it can minimize the damages that the infecting virus does to the human organs [Baron et. al., 1996]. If the antiviral drug can resist the virus from entering the living cell before the virus has the chance to replicate itself, then only the significant damage/ failure of various body organs like lungs, kidney, heart, blood clotting can be avoided. A single antiviral drug can target multiple proteins. In current medical industry we have many examples of such type of antiviral drugs which have multiple use. Similarly, overlapped molecular pathways are also observed for many diseases. For example, Sofosbuvir, Ribavirin and Remdesivir are well known approved drugs for hepatitis C virus [Fried, et al., 2002]. Remdesivir has also established its strong repurposing potential against Ebola, Zika viruses and now CoV-2 virus [Elfiky et al., 2020]. Remdesivir has shown very effective impact on CoV-2 infection by blocking the replication of the virus inside human body [Cao et al., 2020]. Favipiravir and Oseltamivir have been investigated for the treatment of Ebola virus, Lassa virus and influenza viruses A and B. Similarly, Lopinavir and Ritonavir were originally developed for the treatment of HIV patients. Now in current CoV-2 emergency situation all the above-mentioned viral medicines are being used to treat COVID-19 patients [Costanzo, et al., 2020]. Like antiviral drugs, many available antibiotic drugs like Doxycycline, Chloroquine, Hydroxychloroquine etc., also are being used as repurposed drugs for the treatment of various bacterial and viral diseases. Normally antibiotic drugs attacked the walls of a bacteria. Also, these drugs prevent the bacteria from synthesizing a molecule in the cell wall called peptidoglycan. Peptidoglycan provides the wall with the strength it needs to survive in the human body [Kapoor, 2017]. Doxycycline is frequently used for the treatment of various infections by gram-positive and gram-negative bacteria, aerobes and anaerobes and also for other types of bacteria [Grant et al., 2020]. For the treatment of malaria, rheumatoid arthritis, chronic discoid lupus erythematosus [Rainsford, et al., 2015] Hydroxychloroquine is being actively used. For COVID-19 treatment some individual antiviral and antibiotic drugs have already shown lower to moderate effectiveness when tested against infections in patients. Some of them are: Chloroquine, Hydroxychloroquine, Nafamostat, Nitazoxanide, Remdesivir, Ribavirin, Penciclovir, Ritonavir, AAK1, Baricitinib, Choline and Arbidol [Muralidharana, et al., 2020; Chowdhury et al., 2020].

Another effective way of treatment which clinicians are using to fight against COVID-19 is combination drug therapy where two or more different known antiparasitic drugs, immunomodulators or natural remedies are used to treat COVID-19 patients. Combination drugs have already proved their effectiveness to combat against HIV and other coronaviruses earlier [Dyall, et al., 2014; Ter Meulen, et al; 2006; Khan, et al., 2020]. Combination drugs have shown their efficiency for the treatment of MERS-CoV infection [Falzarano, et al., 2013]. Lopinavir and Ritonavir combination is one of the mostly tested repurposed combination medicine for the treatment of MERS-CoV infection [Chan, et al., 2015]. Another combination of drug of Mycophenolic acid (MPA) and IFN-β has proved its strong effectiveness against MERS-CoV affected diseases [Hart, et al., 2014]. In most of the above-mentioned cases, application of combination drugs has been proved to be success as most of the patients were declared clinically recovered [Shalhoub, et al., 2015]. For COVID-19 treatment antibiotic and antiviral drug combination of Hydroxychloroquine and Favipiravir is already being tested [Costanzo et al., 2020]. Similarly, the combination of Hydroxychloroquine + Azithromycin [Gautret, et al., 2020], Favipiravir+ Nafamostat mesylate [Doi et, al. 2020] or Lopinavir+ Oseltamivir + Ritonavir [Muralidharana, et al., 2020] are already in use against SARS-CoV-2 infection.

**Table 1**: Structure of receptor protein (6LU7), Favipiravir (F), Hydroxychloroquine (H) and Oseltamivir (O).

| Compound Name | Structure |
|---|---|
| Protease: 6LU7 | 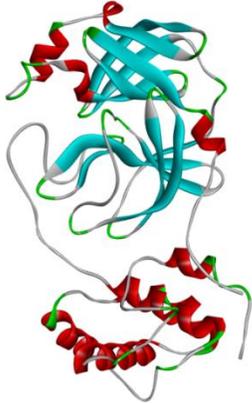 |
| Favipiravir (F) ($C_5H_4FN_3O_2$) | 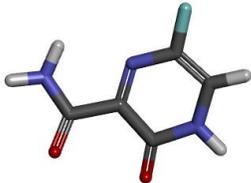 |
| Hydroxychloroquine(H) ($C_{18}H_{26}ClN_3O$) | 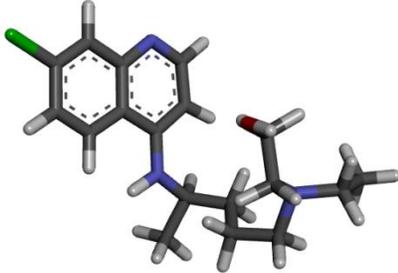 |

| Oseltamivir (O) (C$_{16}$H$_{28}$N$_2$O$_4$) | 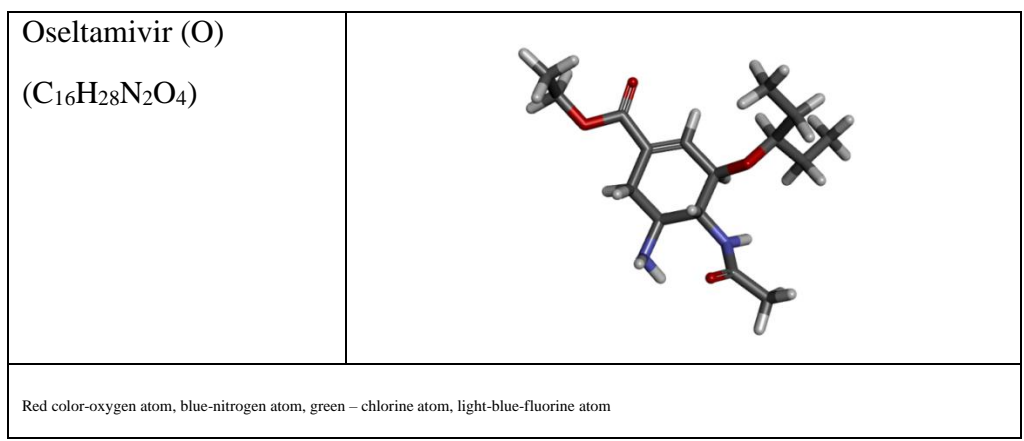 |
|---|---|
| Red color-oxygen atom, blue-nitrogen atom, green – chlorine atom, light-blue-fluorine atom | |

In the present study, we have tried to understand the mechanism of action of some mostly applicable proposed antiviral and antibiotic drug combination for COVID-19. Our proposed combination is Favipiravir +Hydroxychloroquine +Oseltamivir. Favipiravir, a synthetic prodrug has already tested for its antiviral activity against the influenza virus [Furuta, et al., 2002]. It is already been used for the treatment of Ebola virus, Lassa virus, and now COVID-19 [Cai, 2020]. Hydroxychloroquine is an FDA approved drug which is usually being used for the treatment of malaria, rheumatoid arthritis, chronic discoid lupus erythematosus, and systemic lupus erythematosus [Ben-Zvi, et al., 2012]. Oseltamivir is an antiviral drug used for the treatment and prophylaxis of infection with influenza viruses A and B [Hurt, et al., 2009]. In the present work we have represented the challenges and applicability of repurposing of drugs Favipiravir, Hydroxychloroquine and Oseltamivir in terms of their applicability individually and in combination modes against SARS-CoV-2 infections. To understand the interaction between receptor 3CL$^{pro}$ protease and inhibitor ligand drugs we have applied some simulation techniques like energy minimization, molecular docking and molecular dynamics (MD) simulations.

## 2. Materials and Methods:

### *2.1 Procedure of potential target protein structures for SARS-CoV-2*

SARS-CoV-2 is a virus having positively sensed single stranded RNA. The protein structure of CoV-2 contains spike (S) , membrane (M), envelope (E) and nucleocapsid (N) [Woo, et al., 2005; Gorden, et al., 2020]. The structure of CoV-2 virus has been characterized very rapidly after its appearance to the world since the genomic sequence of the virus is already known to the research world [Wu, et al., 2020]. The structures of newly invented CoV-2 virus and already known CoV virus is very similar [Yin, et al., 2020]. The similarity index is about 95%. So the identification process of 3-chymotrypsinlike viral protease (3CL$^{pro}$) of CoV-2 was appeared to be much faster [Jiang et al., 2020].

For quick drug discovery the similarity index plays a major role. To generate the non-structural proteins the 3CL$^{pro}$ split the poly-protein at 11 distinct sites. The process play an significant role in the way of viral replication. 3CL$^{pro}$ is located at the 3 ends, which exhibits excessive variability. Therefore, 3CL$^{pro}$ is a potential target or anti-coronaviruses inhibitors screening [Deng, et al., 2014].

In the present study, we have used 3CL$^{pro}$ proteases(6LU7) as main target protein of drug molecules. 3D structure of the 6LU7 of CoV-2 was retrieved from the Protein Data Bank website (https://www.rcsb.org) [Burley et al., 2019] shown in table 1 and used as a receptor. First of all, from the protein structure the existing water molecules were removed. After that polar hydrogens were added in protein structure. After that, with the help of Discovery studio 2020 the inbuilt ligand was removed from the protein structure [Dassault Systemes BIOVIA, 2017]. All of the above steps were performed in AutoDock [Morris et al., 2009], MGL tools [Tools, 2012]. The output protein structure was saved in PDB format.

## 2.2 Procedure of Ligand drug molecules preparations

There are many drugs available in medical industry, which have several protein targets. Similarly, several diseases are there which share overlapping molecular pathways. So, the concept of drug repurposing has become a very familiar treatment protocol for such type of diseases. In this study, we have used well known and FDA approved drugs Favipiravir ($C_5H_4FN_3O_2$), Hydroxychloroquine ($C_{18}H_{26}ClN_3O$), Oseltamivir ($C_{16}H_{28}N_2O_4$). Favipiravir (T-705) is a synthetic prodrug, which is used in the antiviral activity of chemical agents against the influenza virus [Furuta, et al., 2002]. It has a good bioavailability (~94%), 54% protein binding affinity. Favipiravir-RTP binds to inhibits RNA dependent RNA polymerase (RdRp), which ultimately prevents viral transcription and replication [Furuta, et al., 2017]. Hydroxychloroquine is used in the treatment of malaria, systemic lupus erythematosus [Ben-Zvi, et al., 2012]. The bioavailability of hydroxychloroquine is 67-74%. According to available data the Hydroxychloroquine rises the pH value in human organelles [D'Acquarica, et al., 2020]. The raised pH in human organelles can prevent virus particles (such as SARS-CoV and SARS-CoV-2) from utilizing their activity for fusion and entry into the cell. Oseltamivir is an antiviral drug, which is used for the treatment and prophylaxis of infection with influenza viruses A and B [Hurt, et al., 2009]. Oseltamivir exerts its antiviral activity by inhibiting the activity of the viral neuraminidase enzyme found on the surface of the virus, which prevents budding from the host cell, viral replication, and infectivity [Muralidharana, et al., 2020]. Virtual screening of these drugs have done before the checking of inhibition capability of the F, H and O drugs. SWISS ADME software (https://www.swissadme.ch) and ADMET (https://vnnadmet.bhsai.org/) software were used for the virtual screening of above mentioned drugs [Daina, et al., 2017].

For virtual screening of different available drugs there are some Drug-likeness rules like Lipinski's rule of five (Ro5), Veber's rule, MDDR-like rule, Egan rule, Ghose filter, Muegge rule etc., which are used for preliminary drug screening [Lipinski, 2004, Apparsundaram, et. al., 2000, Veber, et. al., 2002]. Among all of them Ro5 is the most important rule and also known as ''a rule of thumb''. The filters of Ro5 are followed by: Molecular weight less than equals to 500, H-bond donors less than equals to 5, H-bond acceptor less than equals to 10, MLOGP less than equals to 4.15 and molar refractivity between 30 and 140. Those drugs which follows the Ro5 with some required pharmacological properties can be used as a potential candidate for vocally active drug in humans [Lipinski, 2004]. For Drug preparation, the ligands in 'SDF' format were obtained directly from the PubChem (National Library of Medicine) (https://pubchem.ncbi.nlm.nih.gov/) and converted to 'PDB' format with the help of Auto Dock tools [Morris et al., 2009]. After that, all molecular structures of the ligands were optimized by using density functional theory (DFT) with the basis set 6.311G (d,p) [Becke, 1997] using the Gaussian 09 program [Frisch, 2004]. The optimized structure was visualizing with the help of Gauss View 5 molecular visualization program [Dennington, et al., 2007].

**Table 2:** Physiochemical, Drug-likeness, lipophilicity, water solubility and pharmacokinetics properties, Medicinal chemistry and toxicity of Favipiravir, Hydroxychloroquine and Oseltamivir.

| Name of Ligand | Favipiravir | Hydroxychloroquine | Oseltamivir |
|---|---|---|---|
| **Physiochemical Properties** | | | |
| **Molecular Formula** | $C_5H_4FN_3O_2$ | $C_{18}H_{26}C_1N_3O$ | $C_{16}H_{28}N_2O_4$ |
| **Molecular Weight** | 157.10 g/mol | 335.87 g/mol | 312.40 g/mol |
| **Hydrogen Bond Donor Count** | 2 | 2 | 2 |
| **Hydrogen Bond Acceptor Count** | 4 | 3 | 5 |
| **Rotatable Bond Count** | 1 | 9 | 9 |
| **Topological Polar Surface Area** | 88.84 Å² | 48.39 Å² | 90.65 Å² |
| **Heavy Atom Count** | 11 | 23 | 22 |
| **Formal Charge** | 0 | 0.50 | 0.75 |
| **Molar Refractivity** | 32.91 | 98.57 | 84.52 |
| **Lipophilicity** | | | |
| **Log $P_{o/w}$ (iLOGP)** | 0.39 | 3.58 | 2.79 |
| **Log $P_{o/w}$ (XLOGP3)** | -0.56 | 3.58 | 1.10 |
| **Log $P_{o/w}$ (WLOGP)** | -0.57 | 3.59 | 1.29 |
| **Log $P_{o/w}$ (MLOGP)** | -1.30 | 2.35 | 0.63 |
| **Log $P_{o/w}$ (SILICOS-IT)** | 0.69 | 3.73 | 1.33 |
| **Consensus Log $P_{o/w}$** | -0.27 | 3.37 | 1.43 |
| **Water Solubility** | | | |
| **Log $S$ (SILICOS-IT)** | -1.42 | -6.35 | -2.47 |
| **class** | Soluble | Poorly soluble | soluble |
| **Solubility** | 6.04e+00 mg/ml ; 3.85e-02 mol/l | 1.50e-04 mg/ml ; 4.46e-07 mol/l | 1.05e+00 mg/ml ; 3.37e-03 mol/l |
| **Pharmacokinetics** | | | |
| **Gatrointestinal absorption** | High | High | High |
| **BBB permeant** | No | Yes | No |
| **P-gp substrate** | No | No | Yes |
| **CYP1A2 inhibitor** | No | Yes | No |
| **CP2C19 inhibitor** | No | No | No |
| **Log $K_p$ (skin permeation)** | -7.66 cm/s | -5.81 cm/s | -7.42 cm/s |
| **Drug Likeness** | | | |
| **Lipinski Rule** | Yes; 0 violation | Yes; 0 violation | Yes; 0 violation |
| **Ghose Filter** | No; 4 violations: MW<160, WLOGP<-0.4, MR<40, #atoms<20 | Yes | Yes |
| **Veber (GSK) Rule** | Yes | Yes | Yes |
| **Egan (phatmacial) Filter** | Yes | Yes | Yes |
| **Muegge (Bayer) Filter** | No; 1 violation: MW<200 | Yes | Yes |
| **Bioavailability (Abbott) Score** | 0.55 | 0.55 | 0.55 |
| **Medicinal Chemistry** | | | |
| **PAINS (Pan Assey Interference Structures)** | 0 alert | 0 alert | 0 alert |
| **Brenk** | 0 alert | 0 alert | 1 alert: phosphor |
| **Leadlikeness** | No; 1 violation: MW<250 | No; 2 violations: Rotors>7, XLOGP3>3.5 | No; 1 violation: Rotors>7 |
| **Synthetic accessibility** | 2.08 | 2.82 | 4.449 |
| **Toxicity** | | | |
| **Cyto-toxicity** | No | No | No |
| **MRTD(mg/day)** | 170 | 478 | 165 |

## *2.3 Molecular docking and visualization*

To examine the protein-ligand interaction we have used molecular docking using AutoDock 4.2 and AutoDock vina [Morris, et al., 2009]. MGL tools were used for the preparation of protein and ligands for molecular docking. Molecular docking is the computation method which is used to perform the binding energy calculation and energy minimization of the protein- ligand complex. With the help of molecular docking we can identify that the ligand is at their energy minimized state or not with some scoring function. From docking score we can also predict that the ligand is properly docked with host protein (6LU7) or not [Yuriev, et al., 2011]. If the docking of ligand and host protein shows their minimum binding energy, it means this docking result may be showing the feasibility of biochemical reaction. By using AutoDock Vina and Discovery studio visualizer [Dassault Systemes BIOVIA, 2017], the docking based studies and analysis of the recommended inhibitors against protease of COVID-19 through receptor and ligand have been performed. Firstly, with the help of AutoDock tools the structures of prepared protein and ligand were saved in the format of Pdbqt [Morris et al., 2009]. A specific configuration file was used for AutoDock vina algorithm. The parameters for configuration file were as follows: total no. of binding modes is 9, exhaustiveness is -8 and the -kcal/mol energy difference was applied. Grid box was also formed with center x, y, z coordinate of residue position of the receptor protein respectively. For docking based studies, these parameters were used for the recommended inhibitor on the protease of SARS-CoV-2 [Trott, et al., 2010]. After the docking, there were nine different poses (ligand: protein complex structure) obtained. Among all of the poses, the best pose was chosen as such in which, shows maximum non bonded interaction, minimum binding affinity (Kcal/mol), dreiding energy, dipole moment and also minimum inhibition constant. Again, for selecting best pose the maximum no. of hydrogen bond between ligand and receptor protein is another criteria. In this case, firstly, individual of three ligands were docked with receptor protein. After the analysis of individual docking the sequential docking were performed. For sequential docking, the grid box coordinates were established to the specific binding region of individually drug with default grid spacing. The interaction between receptor protein and ligand were analyzed with the help of Discovery Studio Visualizer [Dassault Systemes BIOVIA, 2017]. After that the MD simulation is used for the already docked structure [Berendsen, et al., 1995].

## 2.4 Molecular Dynamic (MD) Simulations

From MD simulation results, we can find out the structural dynamics of protein: ligand interaction at an atomistic level. In MD simulation for calculating thermodynamics parameters of the ligand: protein complex structure the LINUX based platform ''GROMACS 5.1 Package '' [Berendsen et al., 1995] with GROMOS43A2 force fields [Gunsteren, et al., 1996] was used. The thermodynamic parameters computed were : $E_{pot}$, T, D, $R_g$, RMSD, RMSF, SASA, H-bonds and interaction energies. Firstly, all three ligand structures (F, H, O) were optimized by using DFT with the basis set 6.311G (d,p) [Becke, 1997] by using Gaussian 09 package [Frisch, 2004]. All ligands should be at their optimized minimum energy states for molecular dynamics simulation so energy minimization was the first step for MD simulation. For energy minimization of complex, time varying (1ps-100ps) steepest descent algorithm with for 500,00 steps was used. The algorithm has a cut off up to 1000 KJ/mol for minimizing the steric clashes [Anuj, et al., 2020]. There are two phases to obtain the energy minimization of complex with each having 500,000 steps. In the first phase, minimization was obtained through a binary condition from constant particles no. (N), volume(V) and temperature (T) and in second phase particle number, pressure and temperature (NPT) under P=1 atm, T= 300K. Short range Coulomb interaction (Coul-SR) and short range Lennard–Jones interaction (SR-LJ) are used for nonbonded interaction. When all the steps of MD simulation completed, after that the results were analyzed with the help of graphical tool origin. In this study, we have carried out MD simulations for the independent ligands (F, H, O) and also for the combination modes (F+H, F+H+O) in presence of target protein and studied the possibility of formation of complex structures in solvation modes.

## 2.5 Computational details

MD simulations and corresponding energy calculations have been computed using HP Intel Core i5 - 1035G1 CPU and 8 GB of RAM with Intel UHD Graphics and a 512 GB SSD.

## 3. Results and discussion

### 3.1 Analysis of Drug Likeness properties of Favipiravir, Hydroxychloroquine and Oseltamivir

For virtual screening of different available drugs, there are some Drug-likeness rules which have already discussed in the section 2.2. Among all of them rules Ro5 is the most important rule. F, H and O follow the Lipinski's rule of five (Ro5) which can be seen in data already revealed in table 2. All drugs have synthetic accessibility count <5, means all the drugs can synthesized easily. Also, they follow veber rule and so they all satisfy bioavailability conditions. Those drugs which follows the Ro5 rule are proposed as chemical compound which have certain compulsory pharmacological properties which can make the drug as a potential candidate for orally active drug in humans [Lipinski, 2004]. Also RO5 helps to avoid preclinical and clinical failures by primary preclinical progress of drugs by differentiating between nondrug and drug like molecules. From the ADMET analysis we can conclude that none of our tested drug has any cyto-toxicity effect. According to data calculated from ADMET software (https://vnnadmet.bhsai.org/), the maximum suggested dose for F, H and O were obtained as 170mg/day, 478mg/day and 165mg/day.

## *3.2 Analysis of molecular docking results*

Molecular docking of all the three drugs with the receptor protein were performed by Auto dock vina. Discovery Studio Visualizer software [Dassault Systemes BIOVIA, 2017] was used for analyzing the molecular docking results. Firstly, in case of F, for pose 1 we obtained the lowest binding energy (-4.4 kcal/mol), dreiding energy (53.0422) and inhibition constant ($5.9\times10^{-4}$M) at 300K (room temperature) and so pose 1 was considered as the best fit ligand: receptor (F: 6LU7) complex structure (Figure 1a, Table 3). Similarly, in case of H and O, for pose 1 we obtained the lowest binding energies (-4.8 Kcal/mol and -5.1 Kcal/mol), dreiding energies (332.834 and 327.416) and inhibition constants ($3.0\times10^{-4}$M, $1.8\times10^{-4}$M) at 300K (room temperature) for best fit ligand: receptor complex structures (H: 6LU7 and O: 6LU7) (Figure 1b, c and SD1). Lowest value of inhibition constant ($k_i$) validate the strong interaction of ligands (F, H, O) as inhibitor towards the receptor protein 6LU7. To calculate inhibition constant, the equation used was

$$K_i = e^{\Delta G/RT} \dots\dots\dots\dots\dots\dots\dots\dots\dots\dots\dots (1)$$

Where, binding affinity is G, universal constant is R and T temperature (300K)

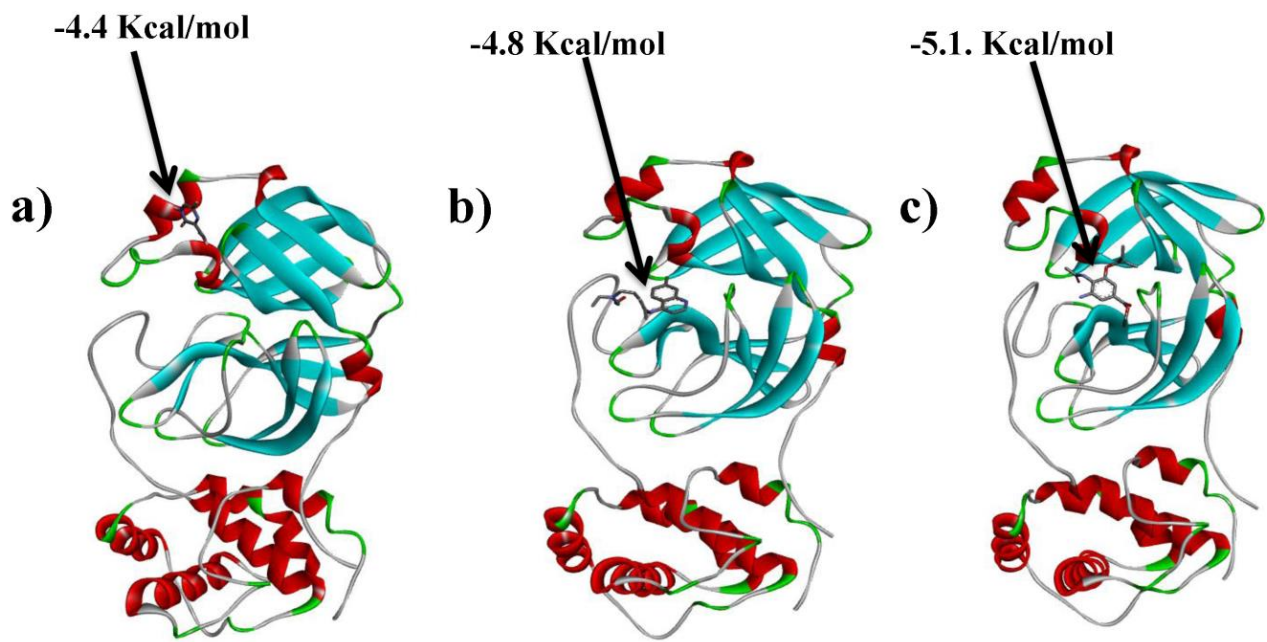

**Figure 1:** Binding energies and binding sites for drugs **a)** F, **b)** H and **c) O** towards protein 6LU7 by individual docking.

The sequential docking mechanism has been used for docking the combination drugs as inhibitors for receptor protein 6LU7. In the present study we have used various combination of drugs. They are: F+H, F+O, H+O and F+H+O. Among the combination of two inhibitors, F+H combination has showed better binding affinity towards 6LU7 from sequential docking. The binding affinity of combination of two drugs F+H: 6LU7 complex showed the substantial enhancement in the binding energy (-5.2 Kcal/mol) from their individual complexation with 6LU7. Similarly, for the three combination drugs F+H+O, further substantial enhancement in the binding affinity (-5.3 Kcal/mol) was observed towards 6LU7. So, from the results of individual and sequential docking we may conclude that the binding affinity for F+H+O combination is better than F+H combination and individual forms of F, H and O. The result of individual docking and sequential docking are shown in figure 2. For every best pose, the donor–acceptor surface are shown in figure 3 in 3D and 2D view with their possible hydrogen bonding (Table 3 and SD1).

**Table 3:** Various interaction parameters for the docked structure of F, F+H and F+H+O with receptor protein 6LU7.

| Ligand | Binding affinity (kcal/mol) | Hydrogen bonded interaction(donor: acceptor, distance in Å)[Type of bond] | Dreiding energy (ligand) | Dipole moment of ligand (Debye) | Inhibition Constant( M) $K_i=e^{\Delta G/RT}$ |
|---|---|---|---|---|---|
| F | -4.4 | (A:THR24:HN -:UNK0:O, 2.00099)[C- H Bond]<br>(A:THR24:HG1 -::UNK0:O,2.30365) [C-H Bond]<br>(A:THR25:HN-: :UNK0:O,2.82701) [C-H Bond]<br>(:UNK0:H -: A:CYS22:O,2.76702) [C-H Bond]<br>(:UNK0:H -: A:THR25:OG1, 2.39306) [C-H Bond]<br>(:UNK0:H -: A:VAL42:O, 2.80565) [C-H Bond]<br>(:UNK0:H -: A:CYS44:O, 2.54365) [C-H Bond]<br>(:UNK0:H -: A:ILE43:O,2.72896) [C-H Bond] | 53.0422 | 2.437 | $5.9 \times 10^{-4}$ |
| F+H | -5.2 | (A:GLN110:HE21-: :UNK0:N,2.70214) [C-H Bond]<br>(A:GLN110:HE2 -: :UNK0:N,2.72991) [C-H Bond]<br>(:UNK0:H -:A:THR111:O,2.5057) [C-H Bond] | 171.402 | 2.444 | $1.5 \times 10^{-4}$ |
| F+H+O | -5.3 | (:UNK0:H30 -:A:LYS137:O, 2.4748) [C-H Bond]<br>(:UNK0:H31 -: A:LYS137:O,2.26105) [C-H Bond]<br>(:UNK0:H32 -:A:GLY138:O, 2.24204) [C-H Bond] | 100.595 | 3.964 | $1.3 \times 10^{-4}$ |
| **C-H Bond- [Conventional Hydrogen Bond]** | | | | | |

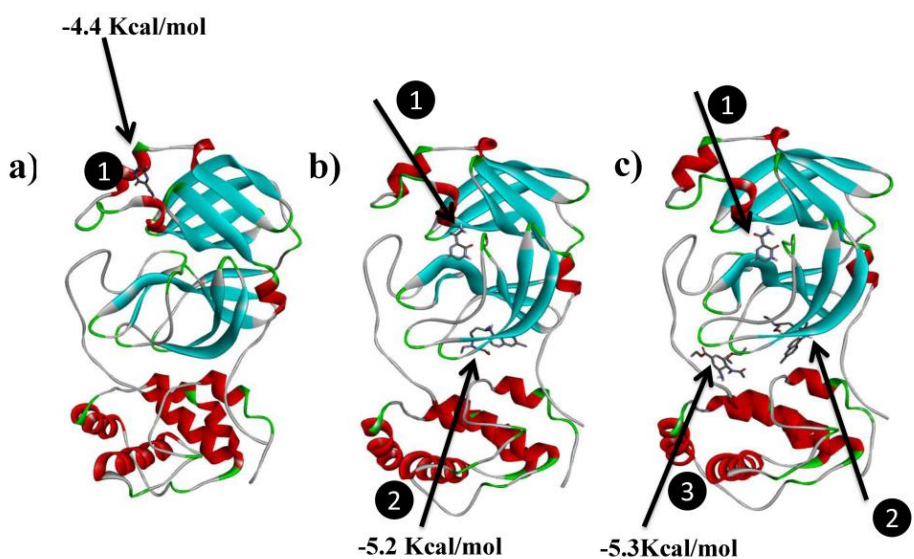

**Figure 2:** Binding energies and binding sites for drug **a)** F, **b)** F+H and **c)** F+H+O towards protein 6LU7 by sequential docking.

The individual protein and ligand shows a different value of dreiding energy. But when the receptor protein and ligand form the complex then they shows the less value of dreiding energy than the individual value. Lowest dreiding energy means the most favorable structure of protein: ligand complex structure [Stephen, et al., 1990].

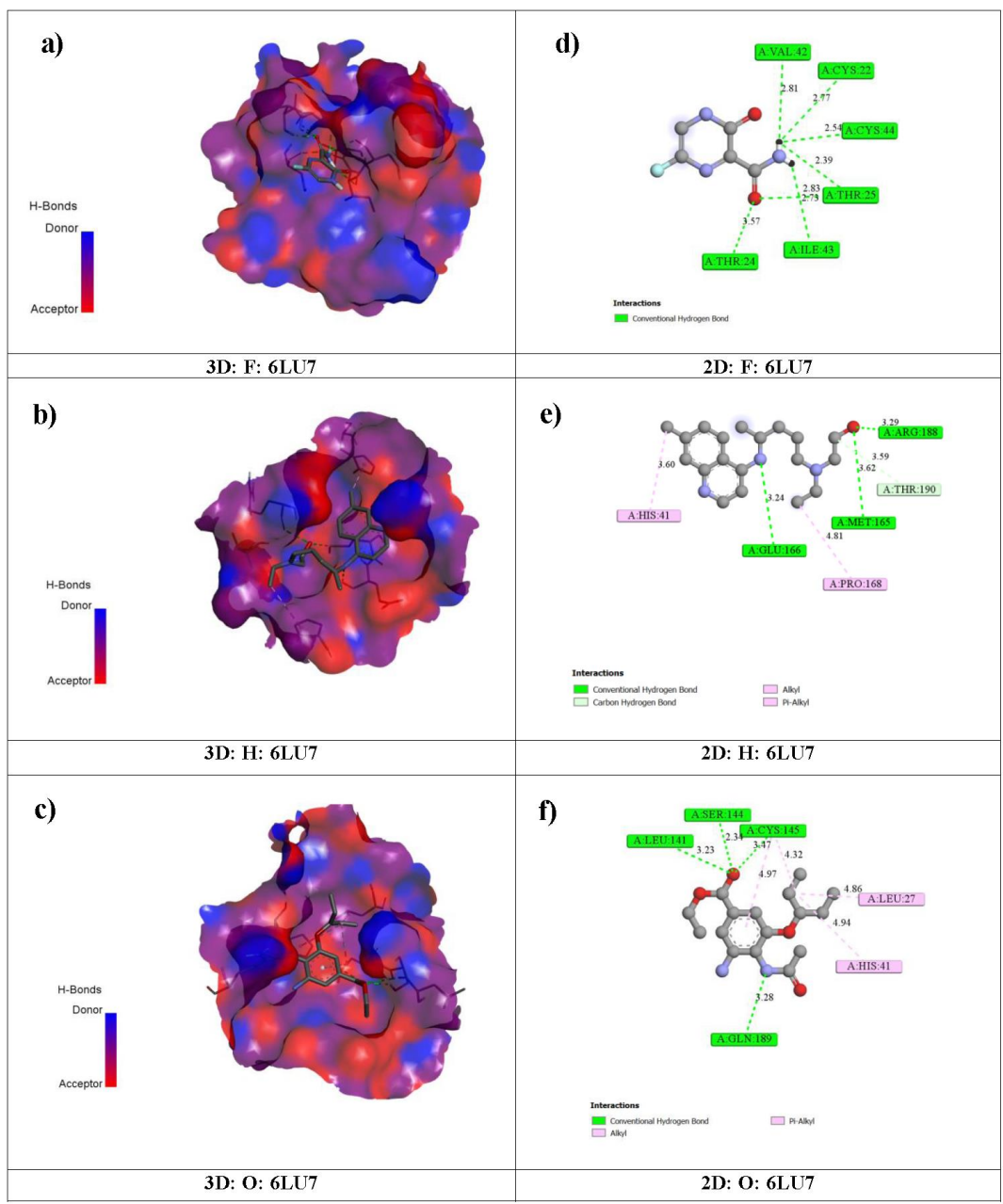

**Figure 3:** Donor: acceptor surface for best pose in terms of H-bond interaction **a), b), c)** F: 6LU7, H: 6LU7, O: 6LU7 **d), e), f)** Possible types of interaction in pose obtained from F: 6LU7, H: 6LU7, O: 6LU7.

In the same way the lowest k$_i$ value for the F+H+O: 6LU7 complex structure means the highest affinity of combination drugs F+H+O towards host protein 6LU7. Results of molecular docking indicates that F+H+O combination can be easily inhibited into the favorable active site of receptor protein 6LU7 and can easily form a best probable stable complex F+H+O: 6LU7. Existence of intermolecular hydrogen bonding between receptor and inhibitor also plays a dominant role for the inhibitor: receptor complex formation. Appearance of hydrogen bonded interactions between F+H+O and 6LU7 protein residues also validated the possibility of complex formation between F+H+O and 6LU7. Through MD simulation, further we check the possibility of existence of various interaction forces needed for the ligand: receptor complex formation between ligand inhibitor and receptor protein for various ligand drugs F, H, O in their individual forms and in their combination modes such as F+H, F+H+O have been discussed through MD simulation results in next section.

### *3.3 MD Simulation analysis*

According to the procedure followed for MD simulation, TIP3P water model has been used and 4Na$^+$ ions were added to maintain the neutrality of ligand: protein complex structure. To calculate the interaction free energies for the protein: ligand complex structure (ΔG$_{bind}$), the MMPBSA (Molecular Mechanics Poisson-Boltzmann Surface Area) method [Rashmi, et al., 2014] sourced from the APBS and GROMACS packages have been used. This model has both repulsive and attractive components [Kollman, et al., 2000]. To ~~predict~~ calculate the binding energy the data was collected at every 100ps between 0ps and 10000ps. In aqueous environment, the binding free energy of the ligand: receptor complex can be shown by following equations.

$$\Delta G_{bind,aqu} = \Delta H - T\Delta S \approx \Delta E_{MM} + \Delta G_{bind,solv} - T\Delta S \ldots \ldots \ldots \ldots \ldots (2)$$

$$\Delta E_{MM} = \Delta E_{covalent} + \Delta E_{electrostatic} + \Delta E_{vander\,waals} \ldots (3)$$

$$\Delta E_{covalent} = \Delta E_{bond} + \Delta E_{angle} + \Delta E_{torsion} \ldots \ldots \ldots \ldots \ldots (4)$$

$$\Delta G_{bind,solv} = \Delta G_{polar} + \Delta G_{nonpolar} \ldots \ldots \ldots \ldots \ldots \ldots \ldots \ldots (5)$$

where, -TΔS, ΔG$_{bind,solv}$ and ΔE$_{MM}$, are the molecular conformational energy due to binding and solvation free energy and mechanical energy changes in gas phase, respectively. In MD simulation, to calculate intermolecular and intramolecular forces of the molecules the force field is the very important factor. From this we can estimate the potential energy for atoms and molecules in the complex. For the force GROMOS43A2 various interaction energies have been computed. The different terms of energies are expressed in additive forms as:

$$E_{total} = E_{bonded} + E_{non\ bonded} \ldots\ldots\ldots\ldots\ldots\ldots\ldots\ldots\ldots. (6)$$

$$E_{bonded} = E_{bond} + E_{angle} + E_{dihedral} \ldots\ldots\ldots\ldots\ldots\ldots\ldots\ldots (7)$$

$$E_{non\ bonded} = E_{hydrogen\ bond} + E_{electrostatic} + E_{vander\ waals} \ldots\ldots (8)$$

$$E_{electrostatic} = E_{coulombic} + E_{lenard\ Jones} \ldots\ldots\ldots\ldots\ldots\ldots\ldots.. (9)$$

To obtain the data at atomistic level MD simulation is used. It can simulate the data in picosecond (ps)/nanosecond(ns) and so it is the best verified silico method among the researchers [McDowell, et al., 2007; Benson & Daggett, 2012]. In this study there are many stable complex structures of drug molecules in their individual levels (F, H, O) and in combinations modes (F+O, F+H, H+O, F+H+O) have been studied which were formed in the molecular docking in presence of main 3CL$^{pro}$ protease (6LU7). MD simulation carried out from 0ps to 10000ps to ~~analyze~~ the stability of forming complex structure. Firstly, in MD simulation all the structures should optimize means every structure has minimum potential energy ($E_{pot}$) and also negative maximum force value. The comparison and analyzation of $E_{pot}$ of the stabilized structures of protein in its bare state with individually docked ligands compound and also with sequential docked ligands have been done simultaneously (Figure 4). Protein in its bare state has an average $E_{pot}$ of $-1.26 \times 10^6 \pm 36.7$ KJ/mol. The average $E_{pot}$ for individual F, H and O and ligands in their combination modes (F+H, F+H+O) in presence of 6LU7 were obtained as at the order of~ $-0.25 \times 10^6 \pm 20.7$ KJ/mol (Figure 4). After optimization we can conclude that all the ligand: 6LU7 complex structures (individual combination) are equally stable and ready for real MD simulation.

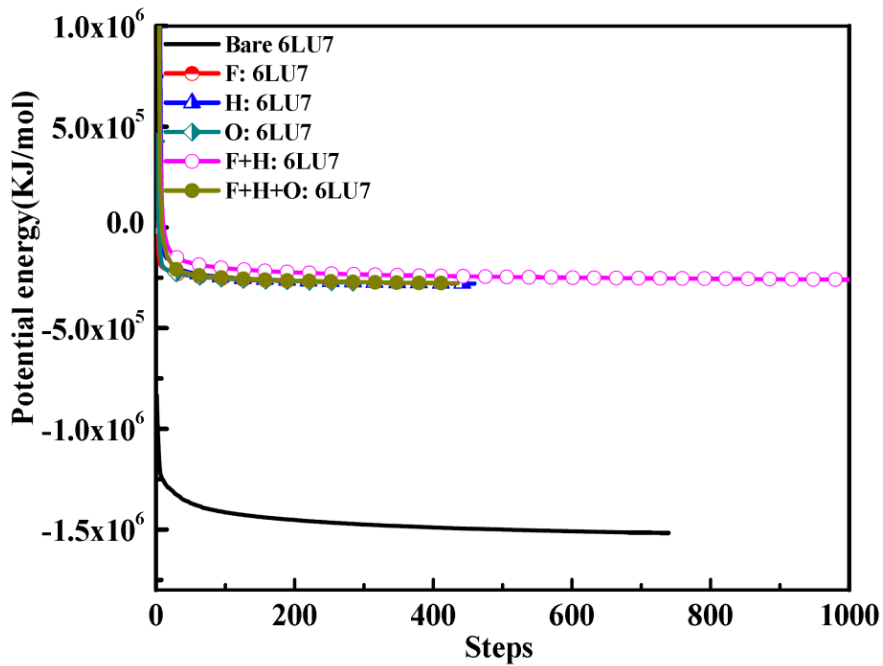

**Figure4:** Potential energy of receptor 6LU7 and complex structures of F, H, O, F+H and F+H+O with 6LU7.

Further, under the equilibrium condition (NVT and NPT) we have verified the stability of each complex structures. The temperature (T), density (D), pressure (P) and volume (V) of each complex structure with varying time trajectory from 0ps to 100ps have been calculated. All computed data for protein in its bare state and ligand: protein complex form showed that the temperature reached to a stable equilibrium state at room temperature (300K) within few ps of simulation (Figure 5a, SD 2a) and sustained the stability through the complete simulation process. The stability of the equilibrated complex structures have been validated by the simulated data of D, P throughout the whole time scale trajectory (Figure 5b, c and SD 2b, c).

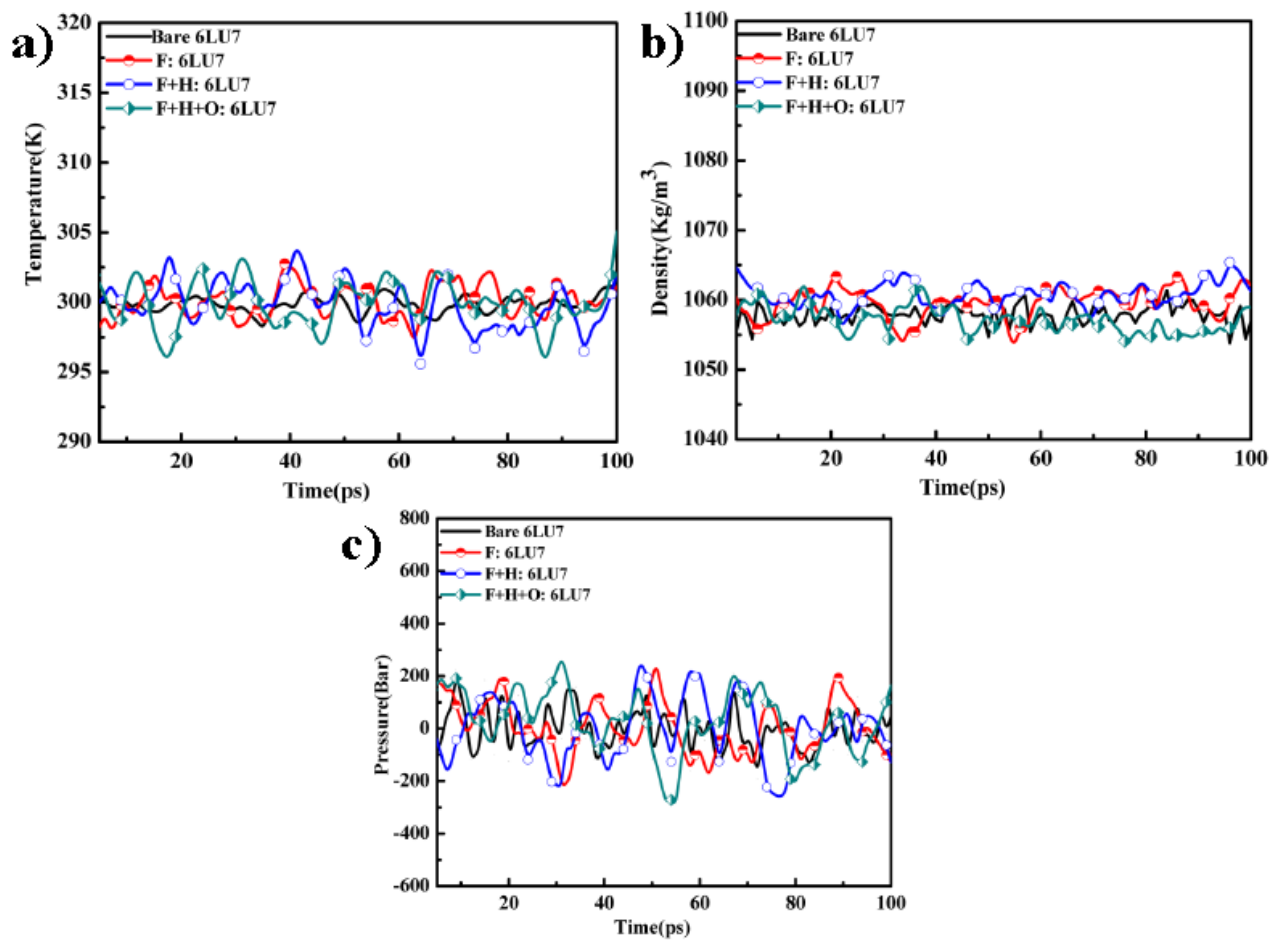

**Figure: 5a)**, **b)**, **c)** Temperature, Density and Pressure computed data for protein 6LU7in its bare state, F:6LU7, F+H: 6LU7, F+H+O: 6LU7 complex.

After the implementation of NVT and NPT conditions the each complex system was completely stable under equilibrium conditions. So system was ready for equilibration process. After running the equilibration process in water medium the MD simulations were operated for complete time trajectory 0-10000ps. Throughout the MD simulation process many thermodynamic parameters for the complex system were computed. These thermodynamic parameters of the complex system are used to know the stability and to observe possible configuration changes of receptor protein in its bare state and complex structure at time resolved simulation trajectory. Some important thermodynamic parameters are RMSD, RMSF, , potential energy, $R_g$, inter-molecular H-bonds, SASA various non-bonded interaction energies for protein: ligand complex structures. Output for all thermodynamic parameters for protein in its bare state and with all complex formations with possible ligands are shown in Table 4, SD 3.

**Table 4:** Statistical data obtained from MD simulations for receptor protein 6LU7 in its bare state and for the F: 6LU7, F+H: 6LU7 F+H+O: 6LU7 complex structure.

| S. No | Parameter | Bare protease (6LU7) | | F: 6LU7 | | F+H: 6LU7 | | F+ H+O: 6LU7 | |
|---|---|---|---|---|---|---|---|---|---|
| | | Mean | Range | Mean | Range | Mean | Range | Mean | Range |
| 1. | SR Columbic Interaction Energy (KJ/mol) | NA | NA | -50±10 | 50--100 | -29.5403±-2.4 | 10--80 | -61.5296±2.9 | 0--80 |
| 2. | SR LJ Interaction Energy (KJ/ mol) | NA | NA | -85±15 | 10- -60 | -195.724±-4.2 | -130--220 | -130.209±3.3 | -30- -70 |
| 3. | RMSD (nm) | 0.22 | 0.13-0.32 | 0.25 | 0.10-0.40 | 0.21 | 0.12-0.29 | 0.22 | 0.14-0.31 |
| 4. | Inter H-Bonds | NA | NA | 5 | 0-9 | 2 | 0-3 | 3 | 0-4 |
| 5. | Radius of gyration | 2.18±0.01 | 2.13-2.24 | 2.21±0.1 | 2.20-2.21 | 2.21±0.2 | 2.16-2.26 | 2.18±0.001 | 2.12-2.25 |
| 6. | SASA ($nm^2$) | 30-35 | 33 | 6-10 | 8 | 6-10.5 | 8.2 | 6-9 | 7.5 |
| 7. | Potential Energy (KJ /mol) | $1.26×10^6±56.7$ | $-7.0×10^5 - -.3×10^6$ | $-0.25×10^6±26.7$ | $-3.4×10^4 - -2.8×10^5$ | $-0.25×10^6±26.7$ | $-3.4×10^4 - -2.8×10^5$ | $-0.25×10^6±26.7$ | $-3.4×10^4 - -2.8×10^5$ |
| 8. | Binding energy(ΔG)(KJ/mol) | NA | NA | -16.871±27.987 | NA | -123.684±67.977 | NA | -214.372±47.627 | NA |
| 9. | Van der Waal Energy($ΔE_{vdw}$) (KJ/mol) | NA | NA | -0.011±0.011 | NA | -146.525±87.807 | NA | -296.393±45.034 | NA |
| 10. | Electrostatic Energy($ΔE_{elec}$)(KJ/mol) | NA | NA | -0.016±0.263 | NA | --8.830±11.978 | NA | -49.616 ±13.931 | NA |
| 11. | Polar Solvation Energy($ΔE_{polar}$)(KJ/mol) | NA | NA | -16.850±27.965 | NA | 45.328±47.037 | NA | 166.828±26.045 | NA |
| 12. | SASA Energy ($ΔE_{apolar}$)(KJ/mol) | NA | NA | -0.006±2.088 | NA | -13.657±67.977 | NA | -27.191 ±4.325 | NA |

In MD simulation, the factor RMSD is very important. It is used to check the stability of complex structure with respect to the (w.r.t) reference carbon backbone structures of receptor protein 6LU7. A 3D view of RMSD values for carbon backbone complex (F+H+O: 6LU7) structure with reference to receptor protein in the whole time trajectory 0-10000ps is shown in figure 6a. Also the 2D graph of RMSD values for carbon backbone for individual complex structures (F: 6LU7, H: 6LU7, O: 6LU7, F+H: 6LU7 F+H+O: 6LU7) with respect to 6LU7 are shown in figure 6b for time trajectory 0-10000ps (Figure 6b, SD 4). From the graph, it is observed for the complex (F+H+O: 6LU7) structure the value of RMSD showed a variation between 0.14-0.31 (±0.01) nm compared to receptor protein 6LU7 variation 0.13-0.32, which indicates that possibility of less fluctuation when the ligands are binding with protein (Table 4). Also the result observed for the complex structures (F: 6LU7, H: 6LU7, O: 6LU7, F+H: 6LU7) showed a variation between 0.10-0.40, 0.23-0.24, 0.09-0.48, 0.12-0.29 nm compared to receptor protein 6LU7 variation 0.13-0.32 again indicates that the less fluctuation in every complex structure during binding with receptor protein (Table 4, SD 3).

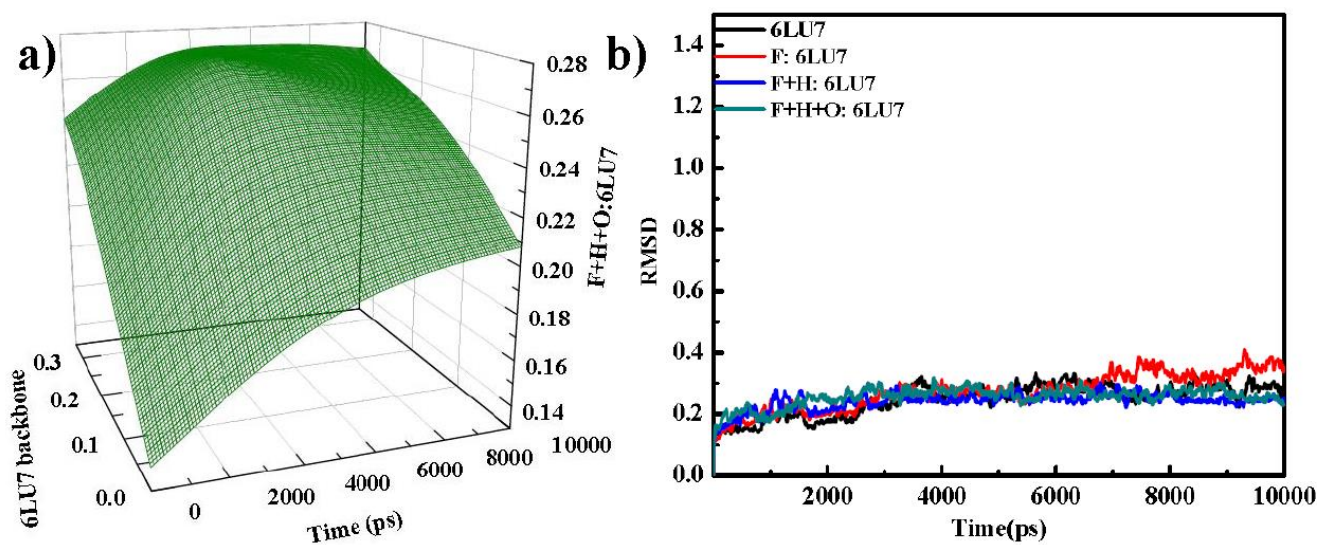

**Figure: 6a)** Root mean square deviation (RMSD) graphs for bare state of receptor protein 6LU7 and in complex (F+H+O: 6LU7) with 6LU7 3D view up to 10000ps and **b)** 2D for complex (F: 6LU7, F+H: 6LU7 and F+H+O: 6LU7) view up to 10000ps.

The average RMSD value for the F+H+O: 6LU7 (0.22nm) and host protein (0.22nm) is same which confirms the best stable complex structure in combination mode. Also the flatness observed in 3D contour suggest that host protein does not show any significant change in presence of F+H+O combination ligand during time resolved simulation which confirmed the better stability of F+H+O: 6LU7 complexation compared to other complexations either in combination modes or in individual forms (Figure 6a). Similarly, in the case of RMSF, among all the complex structures F: 6LU7, H: 6LU7, O: 6LU7, F+H: 6LU7 and F+H+O: 6LU7 we observed less fluctuation between the receptor and inhibitors in all complex structures. Perfect similarity in RMSF values confirmed that the complex structure does not affected the protein backbone (Figure 7a, SD 5a). From all the graph the RMSF value for F+H+O: 6LU7 is exactly same with the host protein means very less fluctuations in the docked structure (Figure7a). The radius of gyration ($R_g$) tells us about the compressed nature of a complex structure or backbone receptor protein [Dharmendra, et al., 2018]. Variation of $R_g$ values throughout the total-time trajectory (0 ps to 10000 ps) showed that all complex structures F: 6LU7, H: 6LU7, O: 6LU7, F+H: 6LU7 and F+H+O: 6LU7 showed a quite stable and compressed structures are shown in table 4. (Figure 7b, SD 5b). $R_g$ of F+H+O: 6LU7 and bare 6LU7 show a perfect match having an average value of 2.18 nm with a fluctuation between 2.13-2.24 nm (Figure7b, Table 4).

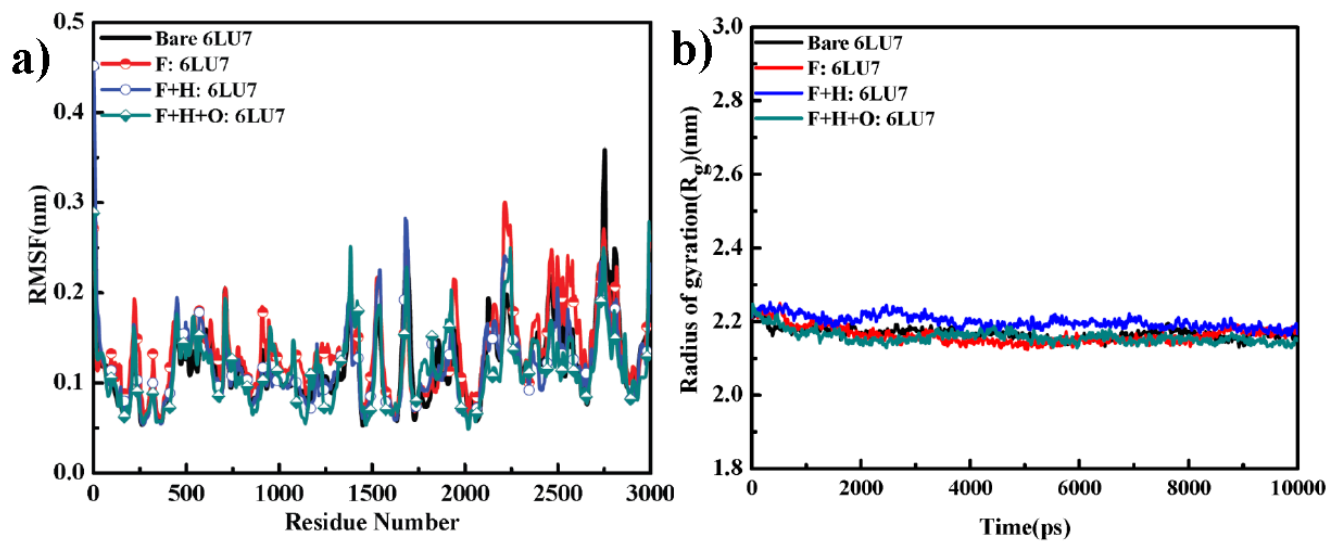

**Figure: 7a)** RMSF for bare state 6LU7 and of complex (F: 6LU7, F+H: 6LU7 and F+H+O: 6LU7) structure, **b)** $R_g$ for bare state 6LU7 and of complex (F: 6LU7, F+H: 6LU7 and F+H+O: 6LU7) structure.

We have also computed solvent Accessible Surface area (SASA) which tells about the area of receptor contact to the solvents. The larger value of SASA means that more of the drug is inserted into the water. And smaller the value of SASA means that more of the drug is covered by the protein means better complexation. The SASA area for bare receptor protein and for all complex structures F: 6LU7, H: 6LU7, O: 6LU7, F+H: 6LU7 and F+H+O: 6LU7 (Figure 8d, SD 6) shown in table 4. The SASA value for bare protein was calculated between 30-35 $nm^2$ with a 33 $nm^2$ mean value however, in presence of different ligand drug molecules, for all inhibitor: 6LU7 complexes the SASA values were observed between 6-10 $nm^2$ which satisfied the possibility of better complexation of all drugs with receptor 6LU7. Among all other drug combination for F+H+O: 6LU7 complex, the lowest value for SASA was observed between 6-9 $nm^2$ with 7.5 $nm^2$ mean value (Figure 8d) which validate the best complexation possibility.

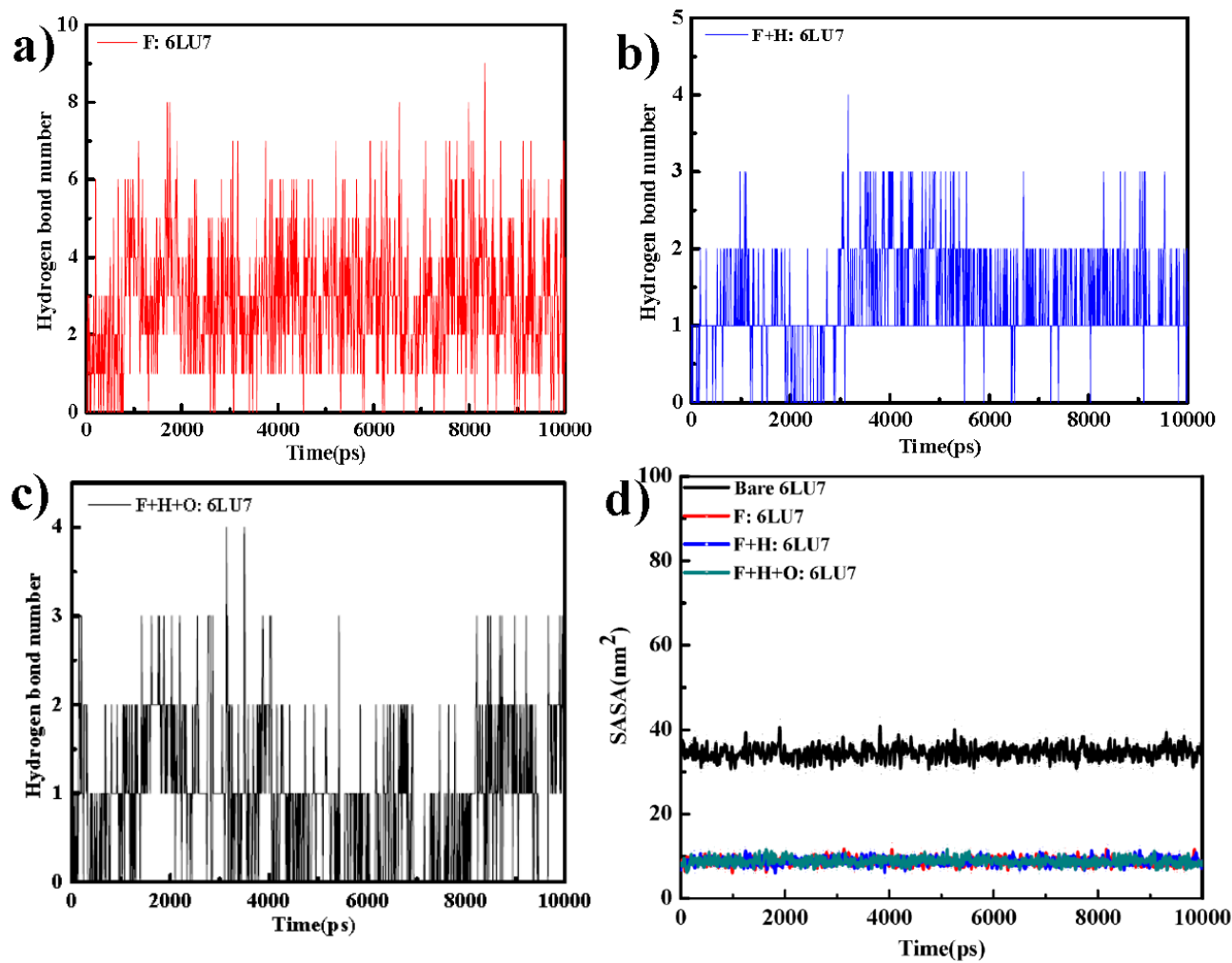

**Figure 8 a), b), c)**: Intermolecular Hydrogen bond numbers for complex (F: 6LU7, F+H: 6LU7, F+H+O: 6LU7) structure for the total time trajectory and **d)** SASA area for protein 6LU7 in its bare state and for complex (F: 6LU7, F+H: 6LU7, F+H+O: 6LU7) structure.

Stability of ligand: receptor protein complex structure is dependent on the contribution of nonbonded interactions. Intermolecular nonbonded hydrogen bonded interaction between ligand and receptor protein plays a dominant role to define the stability of complex structure. For present simulation a 3.5Å cut-off condition is used to identify the proper nonbonded hydrogen bonded interaction. In the present study, the number of hydrogen bonded interactions for different complex combinations F: 6LU7, F+H: 6LU7, F+H+O: 6LU7 were observed to be varying between 0 to 8, 0 to 3 and 0 to 3. The obtained numbers of intermolecular hydrogen bonded interactions between inhibitor: receptor for all combination drugs through MD simulations were perfectly matched with the molecular docking results (Table 4, Figure 8a, b, c).

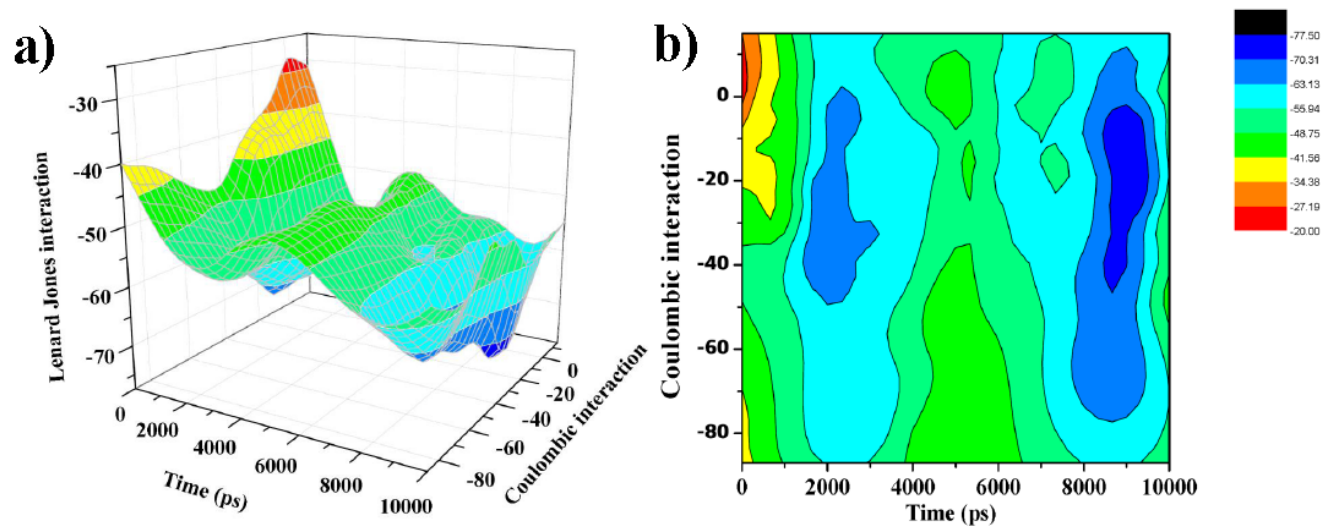

**Figure 9 a):** For F+H+O: 6LU7 complex: **a)**. Variation of Coulombic interaction energy and Lenard Jones interaction energy with respect to time **b)** attached with color contour representation with specific color coding.

Nonbonded interaction energy also plays a very important role to validate the strength of interaction between a ligand and receptor protein in ligand: protein complex structure. In the present study, for all complex structures F: 6LU7, H: 6LU7, O: 6LU7, F+H: 6LU7 and F+H+O: 6LU7, we have observed a variation of short-range Coulombic interaction (Coul-SR) energy and Lennard Jones (LJ-SR) energy (Table 4) over the full-time trajectory (0-10000ps). For all complex structures Lennard-Jones interaction has shown higher effect on the binding affinity than the coulombic interaction. For F+H+O: 6LU7 complexation higher effect of Lennard-Jones interaction energy (-130.209±3.3 KJ/mol) compared to Coulombic interaction (-61.5296±2.9 KJ/mol) is represented by 3D view and with color contour representation of LJ-SR energy and Coul-SR energy with respect to the whole time trajectory (Figure 9b).

To find the presence stronger ligand binding affinities towards receptor protein we have applied the MM/PBSA method. This method deals with Van der Waal energy ($E_{vdw}$), electrostatic energy ($E_{electrostatic}$) and net free binding energy ($\Delta G$). With the help of different quantum simulation techniques, there are a number of research works are going on to check the stability of varieties of complex configurations based on interaction energies [Rana, et al., 2019, Pooja, et al., 2019]. For MM/PBSA method the Van der Waal energy ($E_{vdw}$) components and electrostatic energy ($E_{electrostatic}$) components between inhibitor and receptor are used to determine the stability/binding affinity for all complex structures. The present study is to check the binding affinities of different inhibitor combinations of F, H, O, F+H and F+H+O with receptor protein 6LU7 the above mentioned $E_{vdw}$ and $E_{electrostatic}$ have been computed and are shown in table 4. For nonbonded complexation process between inhibitor and receptor system, $E_{vdw}$ component plays an vital role because it describes the existence of nonbonded interaction in terms of dispersion, repulsion, and induction forces. $E_{electrostatic}$ component may arise due to the unequal distribution of charges between the two components of the complex structure. For ligand: protein complex formation $E_{electrostatic}$ usually does not show any significant role since for ligand: protein nonbonded complex structure there is no major role of charge variation. So greater the value of $E_{vdw}$ means the better stability of complex structure and so the good binding affinity between inhibitor and receptor. Among all inhibitor combinations (F, H, O, F+H and F+H+O), the best binding affinity was observed for F+H+O: 6LU7 with the appearance of maximum value of $E_{vdw}$ (-296.393±45.034kJ/mol) (Table 4). Molecular docking only predicts the binding energy of the protein: ligand complex. The nonbonding interaction energies of the binding region for the complex formation is generally indicated by the ΔG values. ΔG depends on Energy in vacuum, Nonpolar solvation energy and Polar solvation energy for complex structure. The average MM/PBSA free binding energy ($\Delta G_{bind}$) of individual complex structures F: 6LU7, H: 6LU7, O: 6LU7 and F+H: 6LU7 were obtained as -16.871±27.987, -206.849±16.185, -25.973± 33.515 and -123.684±67.977. Whereas the average $\Delta G_{bind}$ for F+H+O: 6LU7 was obtained as -214.372±47.627 kJ/mol (Table 4). The variation of all parameters (energy in vacuum, nonpolar and polar solvation energies) needed for ΔG computation are described in figure 10a, 10b and table 4. Variation of total binding energy for F: 6LU7, H: 6LU7, O: 6LU7, F+H: 6LU7 and F+H+O: 6LU7 over the time trajectory (0-6000 ps) also validate the best binding affinity for the combination inhibitor F+H+O towards the receptor protein 6LU7 by its lowest value throughout the simulation process (Figure 10b). Computed lowest value of $\Delta G_{bind}$ for F+H+O: 6LU7 (-214.372±47.627 kJ/mol) complex structure proved the possibility of best complex formation for the combination drug F+H+O with 3CL$^{pro}$ protease 6LU7. From the results of $\Delta G_{bind}$ we can conclude that combination drug F+H+O strongly binds with receptor 6LU7 protein and shows maximum stability among all other individual and combination drugs.

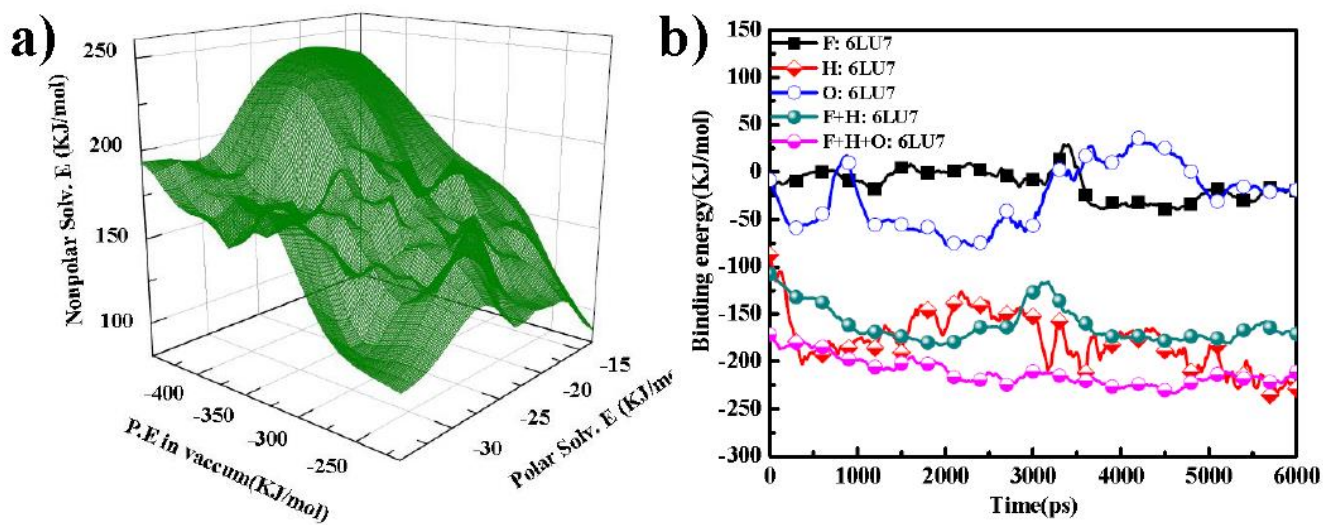

**Figure 10 a):** Relation between energy in vacuum, Polar solvation energy, Nonpolar solvation energy for complex structure F+H+O: 6LU7. **b)** Variation of total binding energy for F, H, O, F+H and F+H+O with receptor protein.

In the present study the MD simulation results validated that repurposing of combination drug F+H+O can make an remarkably stable complex with SARS-CoV-2 protein (6LU7) after binding to the active sites of this protein.

## 4. *Conclusion*

Repurposing of combination drugs have already proved their effectiveness to combat against many viral diseases like HIV, Ebola and against other coronaviruses earlier. For the present study we have tried to recognize the mechanism of action of some combination drugs by repurposing some common antiviral and antibiotic drugs as Favipiravir, Hydroxychloroquine and Oseltamivir against COVID-19. The properties like physiochemical, medical chemistry and pharmacokinetics from ADME analysis have found a strong inhibitory opportunity of F, H, O towards SARS-CoV-2 protein 3CL$^{pro}$. The results of molecular docking validated the stronger binding affinity of combination drug: F+H+O inhibition towards CoV-2 virus (-5.3Kcal/mol) compared to other combination like F+O, F+H or their individual inhibition like F, H, O. Existence of lowest inhibition constant ($1.3 \times 10^{-4}$ M) also established the possibility of better complexation of F+H+O with 6LU7 protease. Different thermodynamical parameters ($E_{pot}$, T, V, D, interaction energies, $R_g$, $\Delta G_{bind}$, SASA energy,) obtained by Molecular dynamics simulations also confirmed the stability of best complex structure between F+H+O combination and CoV-2 protein (6LU7). The perfect closeness of average RMSD between F+H+O:

6LU7 (0.22nm) and host protein (0.22nm) confirmed the total inheritance of proposed combination drug inside host protein. Above result has been further verified by the RMSF variation data which clearly showed that F+H+O: 6LU7 complex structure does not affected the protein backbone. Lowest SASA energy for F+H+O: 6LU7 complex (7.5 nm$^2$) also validated the best stability of F+H+O combination. The stability of F+H+O combination as complex with 6LU7 was further verified by the lowest value of $\Delta G_{bind}$ (-214.372$\pm$47.627 kJ/mol) compared to all other combination and individual forms. By analyzing all the in-silico results obtained from molecular docking and MD simulations we can definitely conclude that repurposing of our proposed antiviral antibiotic drug combination: Favipiravir + Hydroxychloroquine + Oseltamivir has established its strong candidature to be used as a promising potential inhibitor for targeting SARS-CoV-2 virus. We have recommended that our In-Silico results have the strong candidature of combination drugs Favipiravir, Hydroxychloroquine and Oseltamivir as a potential lead inhibitor for targeting SARS-CoV-2 infections.